\documentstyle[12pt,graphics]{article}
\author{Rolf Wittmann \footnote{e--mail: rolf.wittmann@physik.uni-karlsruhe.de} \\
  {\it Laboratorium f\"ur Elektronenmikroskopie, Universit\"at
    Karlsruhe,} \\ {\it Kaiserstr. 12, D--76128 Karlsruhe, Germany}}
\title{Comparing different approaches to model the atomic structure of a ternary decagonal quasicrystal \\ \vspace{0.5cm} {\normalsize (Comparing different decagonal models)}}
\date{\phantom{blabla}}
\newcommand{\qc}{quasicrystal }

\newcommand{\qcs}{quasicrystals }

\newcommand{\deca}{decagonal }

\begin{document}
\maketitle


{\sc Zeitschrift f\"ur Kristallographie, in press}

\begin{abstract}
  It is shown that the covering approach with a single decagonal
  prototile can be transformed into a hexagon, boat and star tiling.
  Particularly, the atomic decoration recently proposed by Cockayne
  and Widom (Phys. Rev. Lett. {\bf 81}, 598 (1998)) as a structure
  model for the \deca phase is investigated. There, conflicts with the
  prototile approach arise which are due to specific peculiarities of the
  decoration.  The above model is compared with recent experimental
  images which give strong support to its main features, but
  contradict competing structure proposals. The implications for the
  stabilization mechanism are discussed.
\end{abstract}

\section{Introduction}

From the beginning, the existence of \qcs has provoked questions about
the physical driving force which results in a highly ordered but very
complicated structure. Especially, the controversy whether truly
deterministic \qcs can exist in real systems or whether \qcs always
contain some inherent random disorder has stimulated much research.
The first approach assumes \qcs to be stabilized energetically, while
the second proposes them to be high--temperature phases due to an
entropic contribution.  

The \qc structure has often been described and analysed in the framework of
quasiperiodic tilings, composed of two or more constituent ``unit
cells''. Recently, a new approach was made by the covering proposed by
Gummelt \cite{Gum96}.  By assuming special overlap rules between
prototiles of decagonal symmetry, it was shown that the resulting
covering of these decagons is equivalent to a deterministic Penrose
tiling. Later, Jeong and Steinhardt \cite{JeS97} gave a simpler proof
and further discussed the physical implications of this cluster
approach. Especially, they showed that the Penrose tiling corresponds
to a maximum density of overlapping decagonal clusters.

The structure of existing \qcs has been analysed experimentally. In
the case of decagonal quasicrystals, structure determinations were
performed e.g. by Steurer et al. \cite{SK90,SHZKL93}. This data was
also the base for some further theoretical structure modeling
\cite{Bur91}. Experimental analyses have been complicated by the
existence of competing structures such as approximants
\cite{HSL91} -- \nocite{ESITS91,GWU94} \cite{GHWW98} or microcrystals \cite{FLDRL94}
which strongly resemble quasicrystals. Furthermore, a problem of all
experimental investigations so far, whether electron microscopy or
x--ray diffraction, is that it is not possible to distinguish between
the different TM atoms in the ternary Al--Co--Cu or Al--Co--Ni \deca
\qcs on which research has focussed.
Concerning the overlap rules of Gummelt, the common experiment--based
models do not provide any crystal chemical motivation.

First attempts to bridge the gap between atomistic structure models
and the mathematical algorithm of Gummelt were made recently
\cite{SJSTAT98,Coc98}. How\-ever, these models were explicitly designed
to impose the Gummelt rules, but without any 
energetic calculations. 

This letter attempts to clarify the relationship of a specific
structure model for a \deca quasicrystal \cite{CoW98}, based on an
atomic decoration of a tiling,
to the prototile framework.  Specific features of the atomic
arrangement prevent a strict equivalency of both approaches and can be
therefore employed to discriminate between them.
Recent experimental data will be discussed and used for comparison
with the theoretical models.

\section{Comparing tiling and covering approaches}

\subsection{Relationship between HBS tiling and prototile covering}

In a first step, the hexagon, boat and star (HBS) tiling \cite{Li95}
will be connected to the overlap rule of decagonal clusters
\cite{Gum96}. For this purpose, the Gummelt decagon (including its
decoration) can be subdivided into a boat and two hexagons. If we now
allow overlaps of the decagons 
of type A (for the nomenclature see ref.~\cite{JeS97}), the following
cases can arise in the HBS framework: i) Two hexagons fall onto each
other -- no problem. ii) A hexagon falls onto a boat -- then the
hexagon should be rejected, leaving only the boat as part of the
tiling. iii) Two boats overlap, forming a star -- then both boats are
to be replaced by a star. (Briefly: if two tiles overlap, they should
be replaced by the tile which encompasses both. Note that larger and
smaller tile always differ by a bowtie shape.) In the case of the type
B overlap, the configuration should be replaced by that shown in
Fig.~1.  Using these rules, we can replace a Gummelt covering by an
HBS tiling.  Checking the allowed surroundings of a decagon (cf.
fig.~14 in ref.~\cite{Gum96}), it can be easily verified that these
rules do not lead to any conflicts or ambiguities.  As each
surrounding in the prototile picture transforms into a different
arrangement of HBS tiles (which always contains a star tile), we can
conclude that these rules even describe a one--to--one correspondence
between Gummelt covering and a HBS tiling. However, as the
covering is equivalent to a Penrose tiling \cite{Gum96}, the HBS
tiling also has to follow specific matching rules.

\subsection{The atomic decoration of Cockayne and Widom}

Recently, Cockayne and Widom (CW) presented a
ternary model for a decagonal quasicrystal  \cite{CoW98} based on Monte--Carlo
simulations and consisting of an HBS tiling supposed to impose
matching rules which would equivalently allow an description in terms
of a Penrose lattice. The atomic arrangement consists of
interconnected decagonal clusters of about 2~nm diameter. Its atomic
decoration shows remarkable similarities to the schematic Gummelt
decoration (Fig.~2). Following the above argumentation, one could
assume the CW model could be equivalently described as a overlap of
decagonal prototiles. In fact, if an HBS tiling is constructed using
the the CW decoration and the above strategy, only a few atoms do not
match when placing a boat on top of a hexagon or a star on top of a
boat. Part of the incongruencies can be resolved by allowing the two
types of TM atoms in a zigzag chain to change places, but not all of
them. In the following, the slight but important differences between
the CW model and the prototile approach will be elucidated.

First, the CW decoration breaks the symmetry both of the boat and the
star tiles. In the case of the boat tile, the asymmetry of positions 1
and 2 in Fig.~2 could be resolved, although this would require also
changes of all other positions marked with a cross. Furthermore, to
symmetricise the boat while allowing a Gummelt overlap of type A would
require not only position 3 but also position 4 to be occupied with a
Cu atom. Together with the Co atom in between, this would result in a
zipper--like interconnection of two zigzag chains of TM atoms. It is
doubtful that such an arrangement would be stable.

An even more ''dramatic'' conflict concerns the bottleneck region of
the bowtie.  There, the two vertices (designated by 5 and 6 in Fig.~2)
approach each other too closely for pentagonal Al clusters to be
centered around both of them.  

Similar difficulties arise also with the decorated star tile.

Therefore, the pentagonal Al clusters at the vertices and
the zigzag chains are stumbling stones to any attempt to construct a
decagonal cluster following the overlap rules exactly. In fact, these
two features are also absent from such atomic models
\cite{SJSTAT98,Coc98}.  In the case of the model of Steinhardt et al.
\cite{SJSTAT98} this is all the more remarkable as part of the authors
already proposed these zigzag chains in a previous paper \cite{STT97}.

The covering forced by the Gummelt decoration is a deterministic
Penrose tiling \cite{Gum96}. Assuming the constituting cluster to be
preferred on energetic reasons, the approach gives a physical
mechanism for producing stable deterministic quasicrystals
\cite{JeS97}. On the other hand, the energetic stability of the
cluster has been only postulated so far. In the specific proposal
\cite{SJSTAT98}, the cluster is built of several hundred atoms, which
makes it difficult to imagine that slightly different clusters will not
be energetically degenerate.

One could argue that such a cluster remains essentially
unaffected by the exchange or displacement of a few of the about
hundred atoms inside. 
Then, the CW model can be considered as a covering of a single
decagonal prototile following the overlap rules of Gummelt, if only
the atomic decoration of the prototile is not strictly fixed but
slight changes and reshufflements are allowed. Therefore, the results
of CW would come close to provide us with a crystal chemical
motivation for the cluster decoration of the Gummelt model. However,
the conceptual beauty of the prototile approach has been lost.

In contrast, the CW model already contains an argument for stability,
namely matching rules induced by ordering of TM atoms. This offers an
explanation at a more fundamental level than having to postulate
special stability of a specific cluster composed of hundreds of atoms.

\section{Comparison of the structure models with experimental results}

In the following, some experimental results will be reviewed and
compared with the above approaches. Tab. 1 contains a summary of the
major differences between the models (including the often--discussed
Burkov model \cite{Bur91}) and their comparison with the experiments.

\subsection{Atomic positions in the 2~nm cluster}

A very peculiar feature of the CW model is the lack of a central atom
in the 2~nm cluster. Instead, the center point is surrounded by
triangular patterns of TM zigzag chains and pentagonal clusters
comprising mainly Al atoms. This contradicts an established consensus
that the 2~nm cluster should have perfect tenfold symmetry which is at
the base of the common structural models
\cite{SK90,Bur91,YaW97}.  High--resolution transmission electron
microscopy (HRTEM) images usually also show a tenfold cluster symmetry
\cite{HLS91,STTTIM96}.  Only in the case of \deca Al--Pd--Mn
\cite{BeH94}, a triangular pattern in the center of the cluster was
reported. How\-ever, it was interpreted to indicate columns
in between the atomic positions, retaining the existence of a central
atom.

On the other hand, HRTEM micrographs contain strong dynamical
contributions and phase contrast due to the microscope lens system.
Therefore, the patterns are not easy to interprete. These
disadvantages can be avoided by using the high--angle annular
dark--field method which directly uses the electrons scattered from
an incident beam
into an annular detector.  The resulting high--resolution images
directly show the atomic structure.
Using this method, images of the decagonal phase with composition
Al$_{72}$Co$_{8}$Ni$_{20}$ have been obtained recently \cite{STTKT97}.
As the scattering power of the Al atoms is five times less than that
of the TM atoms, the latter show up prominently.  Their positions
fit remarkably well to the CW model. Even the triangular arrangement
in the center of a 2~nm cluster column predicted by CW can be clearly
recognized in the experimental images. The HBS tiles of CW are
indicated by broken lines in fig.~3a of ref.~\cite{STTKT97}.  Matching
rule violations are not observed.
The vertices of the tiling show only a weak scattering contrast,
indicating an Al atomic position, which is surrounded by a pentagonal
arrangement of strong scatterers. Saitoh et al. \cite{STTKT97,STT98}
explain this arrangement by a cluster (named S and marked with a star)
motivated by its resemblance to clusters found in the
Al$_{13}$Fe$_{4}$--type approximant. How\-ever, to account for the
experi\-mental observation, the central TM atom had to be arbitrarily
substituted by Al, which gives unrealistic bond lengths. Furthermore,
the tips of the stars indicating clusters of type S always overlap, as
the bright dots forming these pentagons seem to be slightly elongated.
This should not occur with the cluster proposed by Saitoh et al., but
fits perfectly well to the TM zigzag chains in the model of
ref.~\cite{CoW98}. As the two TM atomic positions are only 1.5~\AA \ 
apart, they could not be directly resolved due to the experimental
resolution of about 2~\AA.  

The importance of the experimental findings of Saitoh et al. was
already pointed out by Steinhardt et al. \cite{SJSTAT98}. Although
their model is able to reproduce the observed positions of strong
scatterers quite well by single TM atoms, it does not offer any
explanation for their elongated shape.  It was demonstrated that the
model is also able to reproduce conventional HRTEM reasonably well,
but the same is claimed by part of the coauthors for a different model
\cite{STT98}. Here, the comparison is made with images interpreted
originally to support a cluster model with tenfold symmetry
\cite{HLS91}. These contradictions only reflect the well--known fact
that there is no unique relation between an HRTEM image and a
structure model (even in the case of simple crystalline structures).

Unfortunately, literature data do not
provide a complete set of microscope parameters and imaging conditions
which would be necessary to compare with image simulations of the CW
model. Therefore, a more quantitative comparison of the CW model with
standard HRTEM images has to be left for some future
research.

Meanwhile, another investigation using the same material and the same
method has been published \cite{YPT98}. This time, the resolution of
the microscope was sufficient to resolve the zigzag chains of TM atoms
(called ``buckled columns'' in this reference), thereby giving further
evidence for the correctness of the CW approach and refuting the model
of Steinhardt et al. \cite{SJSTAT98}. According to the study of Yan et
al. \cite{YPT98}, the material is composed from two types of clusters,
one with the characteristic triangular pattern in the center, the
other where the decagonal symmetry is retained even in the central
part. The latter would contradict both the CW and the Steinhardt et
al.~model. How\-ever, it should not be forgotten that all HRTEM images
inherently project all atomic positions along the direction of
observation. The second type of cluster can be easily explained by
assuming clusters of the first type, but stacked on top of each other
rotated by multiples of 36$^{\circ}$. What has been called chemical
disorder \cite{YPT98} could also be interpreted as stacking disorder.
These rotations, if acting in an ordered way, could also explain
superstructures along the decagonal axis. Indeed, higher periods of
0.8, 1.2 and 1.6 nm were experimentally observed \cite{HWK88}. If
acting in a random way, they constitute a source of entropy. As every
structure determination, whether by HRTEM or by Patterson analysis,
contains an averaging, it also can be suggested why the majority of
models contain a cluster of perfect decagonal symmetry
\cite{SK90,Bur91,YaW97}.

\subsection{Further comparison}

So far, the comparison was focussed only on the atomic positions
projected along the decagonal axis. In the following, other features
will be checked. 

The experimentally investigated phase contains a $10_{5}$ screw axis
which is correctly reproduced by both models of CW and Steinhardt et
al.

Steinhardt et al. \cite{SJSTAT98} calculate a composition of
Al$_{72.5}$Co$_{8.9}$Ni$_{18.6}$ for their model, which is equal to
the experimental composition within the error margins.

Starting from the composition Al$_{64.8}$Co$_{19.6}$Cu$_{15.6}$ of the
CW model \cite{CoW98} and bearing in mind that Cu substitutes for Al
and Co in equal proportions, we obtain a pseudo--binary
Al$_{72.6}$TM$_{27.4}$ phase, which also compares remarkably well with the
composition Al$_{72}$TM$_{28}$ of the experimental studies
\cite{STTKT97,YPT98}. How\-ever, the Monte--Carlo simulations of binary
Al--Co would suggest a quite different structure for the decagonal
phase of such a composition \cite{CoWi98}. The \qc of Saitoh et al.
also does not correspond to the basic Co--rich but to the basic
Ni--rich phase. Both variants are connected by several superstructures
along the Al$_{72}$Co$_{x}$Ni$_{28-x}$--line \cite{RBNGSL98}. As Co
and Ni would be treated similarly by the Monte Carlo approach, one
would not expect any variations with $x$.  Therefore, the r\^ole of
transition metals seems to be more subtle.
It should be admitted, however, that there remains a major aspect of
the CW model which still awaits experimental verification, namely the
alternation of the two different TM atoms along a zigzag chain. As Co
and Cu or Ni cannot be distinguished by electron microscopy or
standard x--ray techniques, this test will be rather difficult to
perform.

Some caution should be applied in generalizing the results obtained on
the above composition to ``the'' decagonal phase. It is by now
commonly accepted that even in one alloy system different variants
exist \cite{RBNGSL98}. Their structures might be more different than
previously assumed. In this respect it is interesting to recall the
neutron scattering experiments performed on
Al$_{72.5}$Co$_{11}$Ni$_{16.5}$ exhibiting superstructure reflections
which, however, were concluded not to be due to a Co/Ni ordering but a
true structural feature \cite{HFG97}.

\subsection{Thermodynamical considerations}

Experimentally, the basic--Ni decagonal structure was observed to be
stable only as a high--temperature phase \cite{RBNGSL98}, indicating
an entropic contribution which stabilizes the structure. On the other
hand, the prototile approach was advanced mainly to provide a
plausible mechanism for energetic stabilization via maximizing the
density of clusters assumed to have minimum energy \cite{JeS97}.
Therefore, the prototile idea lacks experimental support also from the
thermodynamic viewpoint. Although  the CW model was also derived by
energetic arguments,  it might be more flexible to accomodate
random components.

The concept of random tilings \cite{Hen91} is able to provide an entropic term for stability. 
However, a recent tiling analysis \cite{JRB97} showed the
high--temperature basic Ni--rich phase to be a perfect Penrose tiling.
(This contrasts to a variant with superlattice ordering where even
single phasons could be identified by HRTEM \cite{RNB96}.) The evident
entropic term was attributed to chemical disorder \cite{YPT98,JRB97}
without specifying its exact origin. 

In the following some mechanisms will be listed which can introduce
disorder easily and in a into the CW model without changing the
underlying tiling: i) Switching between the two mirror--related
variants of the boat tile, or rotating the star tile which also
results in non--equivalent atomic configurations due to the symmetry
breaking of the decoration; ii) rotating the 2~nm cluster, as
mentioned above; iii) exchange of Co and Cu in the zigzag chains.

\section{Summary}

An algorithm is presented which transforms a covering with decagonal
prototiles and the overlap rules of Gummelt into an HBS tiling. The
atomic decoration of an HBS tiling proposed Cockayne and Widom,
although it resembles the Gummelt decoration very much, contains
features which do not allow it to be interpreted as a reasonable
realization of such a covering in the strict sense. Only if minor
variations in the constituting decagonal cluster are allowed, the
approaches could be reconciled.  Experimental investigations render
support to the major features of the model of Cockayne and Widom, but
contradict the competing proposal of Steinhardt et al. 
Tab. 1 contains a brief overview of the comparison. Chemical
disorder has been proposed as a means to obtain an entropic
stabilization which is suggested by the experimentally determined
phase diagram.

\section*{Acknowledgment}

The author gratefully thanks Eric Cockayne for fruitful discussions
and valuable comments.
The project was supported by the Deutsche Forschungsgemeinschaft
(grant no.~Wi1645/1).

\section*{Table}

\hspace{-2cm}\begin{small}
\begin{tabular}{||l||l|l|l||l|l||}
\hline
& Cockayne and   & Steinhardt  & Burkov \cite{Bur91} & Saitoh & Yan  \\
& Widom (CW)\cite{CoW98} & et al. \cite{SJSTAT98} & & et al. \cite{SJSTAT98}  & et al. \cite{YPT98} \\
\hline
\hline 
construction & HBS tiling & covering & covering / & & \\
& & & tiling & & \\
\hline
central atom & no & no & yes & no & no \\
\hline
tenfold symmetry & & & & & sometimes \\
in cluster center & broken  & broken & perfect & broken & broken \\
\hline 
zigzag chains & & & & & \\
of TM atoms & yes & no & no & not resolved & yes \\
\hline
Co/Cu ordering & yes & yes & no & ? & ? \\
\hline    
screw axis & $10_{5}$ & $10_{5}$ & $10_{5}$ & $10_{5}$ & $10_{5}$ \\
\hline
composition & Al$_{64.8}$Co$_{19.6}$Cu$_{15.6}$ & Al$_{72.5}$Co$_{8.9}$Ni$_{18.6}$ & Al$_{60}$TM$_{40}$ & Al$_{72}$Co$_{8}$Ni$_{20}$ & Al$_{72}$Co$_{8}$Ni$_{20}$  \\
            & =  Al$_{72.6}$TM$_{27.4}$ & & & & \\
\hline
stabilisation & energetic & energetic & & (entropic \cite{RBNGSL98}) & (entropic \cite{RBNGSL98}) \\
\hline
\end{tabular}
\end{small}

\setlength{\parindent}{0pt}Tab.~1: \\
Comparison of special features of the different structure models (left
part), especially those suitable for comparison with the experimental
results (right part).

\section*{Figure captions}

Fig. 1: \\ 
Part of an HBS tiling corresponding to a type B overlap of
two decagonal clusters. The boundaries of the original decagonal
clusters are indicated by broken lines.

\vspace{1cm} 

\setlength{\parindent}{0pt}Fig.~2: \\
Atomic cluster of 2 nm diameter taken from ref.~\cite{CoW98} (fig.~2)
and formed by a boat and two hexagons of the CW model.  The Al atoms
at the vertices of the HBS tiling and the few Al atoms marked by a
cross are located at $z = 0 \mbox{ or } 0.5$ whereas all other atoms
occupy positions at $z = 0.25 \mbox{ or } 0.75$ (in units of the
$c$--axis periodicity).  The cluster is superposed on the Gummelt
decoration (indicated by thick lines) of a decagonal prototile,
showing large similarities. The relevance of the atoms designed by
numbers 1 to 6 is discussed in the text.

\clearpage \newpage
 
\includegraphics{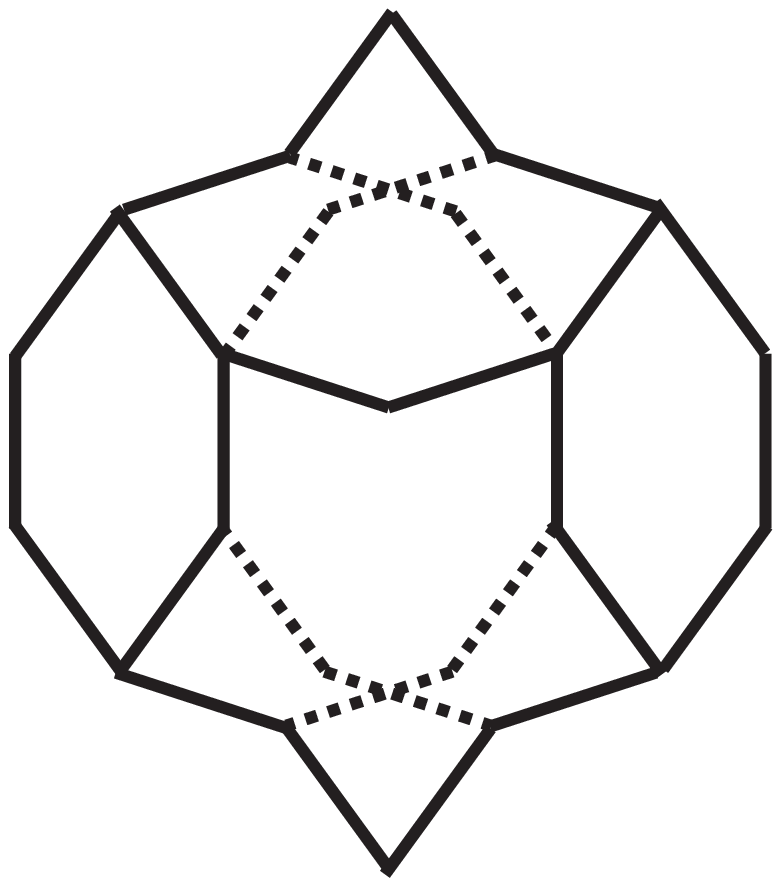}

\vfill

{\tt Fig. 1}

\clearpage \newpage

\includegraphics{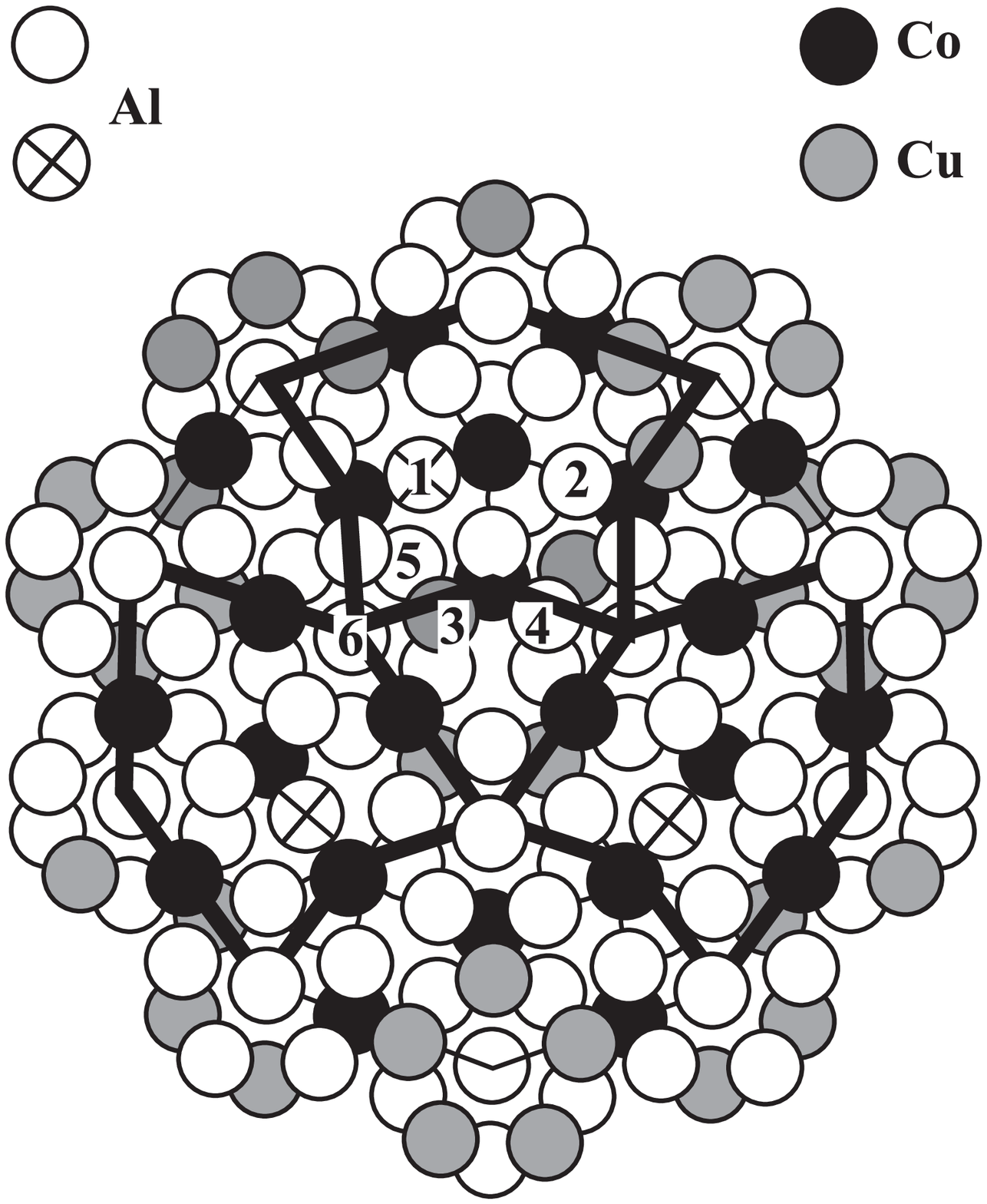}

\vfill

{\tt Fig. 2}


\begin{thebibliography}{10}

\bibitem{Gum96}
P.~Gummelt.
\newblock Penrose tilings as coverings of congruent decagons.
\newblock {\em Geometriae Dedicata}, {\bf 62}:1 -- 17, (1996).

\bibitem{JeS97}
H.-C. Jeong and P.J. Steinhardt.
\newblock Constructing {P}enrose--like tilings from a single prototile and the
  implications for quasicrystals.
\newblock {\em Phys. Rev. B}, {\bf 55}:3522 -- 3532, (1997).

\bibitem{SK90}
W.~Steurer and K.H. Kuo.
\newblock Five--dimensional structure analysis of decagonal
  {A}l$_{65}${C}u$_{20}${C}o$_{15}$.
\newblock {\em Acta Cryst.~B}, {\bf 46}:703 -- 712, (1990).

\bibitem{SHZKL93}
W.~Steurer, T.~Haibach, B.~Zhang, S.~Kek, and R.~L\"uck.
\newblock The structure of decagonal {A}l$_{70}${N}i$_{15}${C}o$_{15}$.
\newblock {\em Acta Cryst.~B}, {\bf 49}:661 -- 675, (1993).

\bibitem{Bur91}
S.E. Burkov.
\newblock Structure model of the {Al--Cu--Co} decagonal quasicrystal.
\newblock {\em Phys. Rev. Lett.}, {\bf 67}:614 -- 617, (1991).

\bibitem{HSL91}
K.~Hiraga, W.~Sun, and F.J. Lincoln.
\newblock Structural change of {A}l--{C}u--{C}o decagonal quasicrystal studied
  by high--resolution electron microscopy.
\newblock {\em Jap.~J.~Appl.~Phys.}, {\bf 30}:L 302 -- L 305, (1991).

\bibitem{ESITS91}
K.~Edagawa, K.~Suzuki, M.~Ichihara, S.~Takeuchi, and T.~Shibuya.
\newblock High--order periodic approximants of a decagonal quasicrystal in
  {A}l$_{70}$--{N}i$_{15}$--{C}o$_{15}$ alloy.
\newblock {\em Phil.~Mag.~B}, {\bf 64}:629 -- 638, (1991).

\bibitem{GWU94}
B.~Grushko, R.~Wittmann, and K.~Urban.
\newblock Structural variations and transformation behaviour of the
  {A}l$_{68}${C}u$_{11}${C}o$_{21}$ decagonal phase.
\newblock {\em J.~Mater.~Res.}, {\bf 9}:2899 -- 2906, (1994).

\bibitem{GHWW98}
B.~Grushko, D.~Holland-Moritz, R.~Wittmann, and G.~Wilde.
\newblock Transition between periodic and quasiperiodic structures in
  {Al--Ni--Co}.
\newblock {\em J. Alloys Comp.}, {\bf 280}:215 -- 230, (1998).

\bibitem{FLDRL94}
M.~Fettweis, P.~Launois, F.~D\'enoyer, R.~Reich, and M.~Lambert.
\newblock Decagonal quasicrystalline or microcrystalline structure: The
  specific case of {Al--Cu--Co(--Si)}.
\newblock {\em Phys. Rev. B}, {\bf 49}:15573 -- 15587, (1994).

\bibitem{SJSTAT98}
P.J. Steinhardt, H.-C. Jeong, K.~Saitoh, M.~Tanaka, E.~Abe, and A.P. Tsai.
\newblock Experimental verification of the quasi--unit--cell model of
  quasicrystal structure.
\newblock {\em Nature}, {\bf 396}:55 -- 57, (1998).

\bibitem{Coc98}
E.~Cockayne.
\newblock Generation of quasicrystals via a single cluster.
\newblock (unpublished).

\bibitem{CoW98}
E.~Cockayne and M.~Widom.
\newblock Ternary model of an {Al--Cu--Co} decagonal quasicrystal.
\newblock {\em Phys. Rev. Lett.}, {\bf 81}:598 -- 601, (1998).

\bibitem{Li95}
X.Z. Li.
\newblock Structure of {Al--Mn} decagonal quasicrystal. {I}. {A} unit--cell
  approach.
\newblock {\em Acta Cryst. B}, {\bf 51}:265 -- 270, (1995).

\bibitem{STT97}
K.~Saitoh, K.~Tsuda, and M.~Tanaka.
\newblock Structural models for decagonal quasicrystals with pentagonal
  atom--cluster columns.
\newblock {\em Phil. Mag. A}, {\bf 76}:135 -- 150, (1997).

\bibitem{YaW97}
A.~Yamamoto and S.~Weber.
\newblock Five--dimensional superstructure of decagonal {Al--Ni--Co}
  quasicrystals.
\newblock {\em Phys. Rev.~Lett.}, {\bf 78}:4430 -- 4433, (1997).

\bibitem{HLS91}
K.~Hiraga, F.J. Lincoln, and W.~Sun.
\newblock Structure and structural change of {Al}--{Ni}--{Co} decagonal
  quasicrystal by high--resolution electron microscopy.
\newblock {\em Mater. Trans. JIM}, {\bf 32}:308 -- 314, (1991).

\bibitem{STTTIM96}
K.~Saitoh, K.~Tsuda, M.~Tanaka, A.P. Tsai, A.~Inoue, and T.~Masumoto.
\newblock Electron microscope study of the symmetry of the basic atom cluster
  and a structural change of decagonal quasicrystals of {Al--Cu--Co} alloys.
\newblock {\em Phil. Mag. A}, {\bf 73}:387 -- 398, (1996).

\bibitem{BeH94}
C.~Beeli and S.~Horiuchi.
\newblock The structure and its reconstruction in the decagonal
  {Al}$_{70}${Mn}$_{17}${Pd}$_{13}$ quasicrystal.
\newblock {\em Phil. Mag. A}, {\bf 70}:215 -- 240, (1994).

\bibitem{STTKT97}
K.~Saitoh, K.~Tsuda, M.~Tanaka, K.~Kaneko, and A.P. Tsai.
\newblock Structural study of an {Al$_{72}$Ni$_{20}$Co$_{8}$} decagonal
  quasicrystal using the high--angle annular dark--field method.
\newblock {\em Jpn.~J.~Appl. Phys.}, {\bf 36}:L1400 -- L1402, (1997).

\bibitem{STT98}
K.~Saitoh, K.~Tsuda, and M.~Tanaka.
\newblock New structural model of an {Al$_{72}$Ni$_{20}$Co$_{8}$} decagonal
  quasicrystal.
\newblock {\em J. Phys. Soc. Jpn.}, {\bf 67}:2578 -- 2581, (1998).

\bibitem{YPT98}
Y.~Yan, S.J. Pennycook, and A.P. Tsai.
\newblock Direct imaging of local chemical disorder and columnar vacancies in
  ideal decagonal {Al--Ni--Co} quasicrystals.
\newblock {\em Phys. Rev. Lett.}, {\bf 81}:5145 -- 5148, (1998).

\bibitem{HWK88}
L.X. He, Y.K. Wu, and K.H. Kuo.
\newblock Decagonal quasicrystals with different periodicities along the
  tenfold axis in rapidly solidified {A}l$_{65}${C}u$_{20}${M}$_{15}$ ({M} =
  {M}n, {F}e, {C}o or {N}i).
\newblock {\em J.~Mat.~Sci.~Lett.}, {\bf 7}:1284 -- 1286, (1988).

\bibitem{CoWi98}
E.~Cockayne and M.~Widom.
\newblock Structure and phason energetics of {Al--Co} decagonal phases.
\newblock {\em Phil Mag. A}, {\bf 77}:593 -- 619, (1998).

\bibitem{RBNGSL98}
S.~Ritsch, C.~Beeli, H.-U. Nissen, T.~G\"odecke, M.~Scheffer, and R.~L\"uck.
\newblock The existence regions of structural modifications in decagonal
  {Al}--{Co}--{Ni}.
\newblock {\em Phil. Mag. Lett.}, {\bf 78}:67 -- 75, (1998).

\bibitem{HFG97}
K.~Hradil, F.~Frey, and B.~Grushko.
\newblock Single crystal neutron diffraction of decagonal
  {Al}$_{72.5}${Co}$_{11}${Ni}$_{16.5}$.
\newblock {\em Z.~Kristallogr.}, {\bf 212}:89 -- 94, (1997).

\bibitem{Hen91}
C.L. Henley.
\newblock Random tiling models.
\newblock In D.P. DiVincenzo and P.J. Steinhardt, editors, {\em Quasicrystals.
  The State of the Art.}, pages 429 -- 524. World Scientific, Singapore, New
  Jersey, London, Hong Kong, 1991.

\bibitem{JRB97}
D.~Joseph, S.~Ritsch, and C.~Beeli.
\newblock How to distinguish quasiperiodic from random order in {HRTEM} images.
\newblock {\em Phys. Rev. B}, {\bf 55}:8175 -- 8183, (1997).

\bibitem{RNB96}
S.~Ritsch, H.-U. Nissen, and C.~Beeli.
\newblock Phason related stacking disorder in decagonal {Al}--{Co}--{Ni}.
\newblock {\em Phys. Rev. Lett.}, {\bf 76}:2507 -- 2510, (1996).

\end{thebibliography}
\end{document}